\documentclass{iau}
\usepackage{graphicx}
\usepackage{bm}
\usepackage{hyperref}
\hypersetup{
    colorlinks=true,
    linkcolor=red,
    citecolor=blue,
} 
\usepackage{pifont}
\newcommand{\cmark}{\ding{51}}%
\newcommand{\xmark}{\ding{55}}%
\newcommand{\sigp}{{\sigma_{\!  p}}}
\newcommand{\sigx}{{\sigma_{\!  x}}}

\renewcommand{\d}[0]{{\rm{d}}}
\newcommand{\del}[0]{\partial }

\renewcommand{\v}[1]{\bm{#1} }\newcommand{\vx}[0]{\bm{x} }

\newcommand{\vp}[0]{\bm{p} }
\newcommand{\vnabla}[0]{\bm{\nabla}} 

\title[CU 14.~~Beyond single-stream with the Schr\"odinger method] 
{Beyond single-stream\\with the Schr\"odinger method}

\author[C. Uhlemann and M. Kopp]   
{Cora Uhlemann and Michael Kopp}

\affiliation{Excellence Cluster Universe, Boltzmannstr. 2, D-85748 Garching\\
Arnold Sommerfeld Center for Theoretical Physics, LMU, Theresienstr. 37, D-80333 Munich} 

\pubyear{2014}
\volume{308}  
\pagerange{???}
\jname{The Zeldovich Universe: Genesis and Growth of the Cosmic Web}
\editors{R. van de Weygaert, S. Shandarin, E. Saar \& J. Einasto, eds.}
\begin{document}

\maketitle

\begin{abstract}
We investigate large scale structure formation of collisionless dark matter in the phase space description based on the Vlasov-Poisson equation. We present the Schr\"odinger method, originally proposed by \cite{WK93} as numerical technique based on the Schr\"odinger Poisson equation, as an analytical tool which is superior to the common standard pressureless fluid model. Whereas the dust model fails and develops singularities at shell crossing the Schr\"odinger method encompasses multi-streaming and even virialization. 
\end{abstract}


{\it \textbf{Introduction.}}
The standard model of large-scale structure and halo formation is based on collisionless cold dark matter (CDM), a particle species that for this purpose can be assumed to interact only gravitationally and to be cold or initially single-streaming. We are therefore interested in the dynamics of a large collection of identical point particles that via gravitational instability evolve from initially small density perturbations into eventually bound structures, like halos that are distributed along the cosmic web. \\
\vspace{-1.5cm}\\
\section{Phase space description of cold dark matter}
  
 The dynamics of CDM with mass $m$ is described by the one-particle phase space density $f(\v{x},\v{p},t)$ which fulfills the Vlasov-Poisson equation
\begin{equation}
\label{VlasovEq}
\partial_t f=   -\frac{\vp}{a^2 m}\cdot\vnabla_{\! \!  x} f + m \vnabla_{\! \!  x} V \cdot\vnabla_{\! \!  p} f \quad , \quad
\Delta V = \frac{4\pi G\,\rho_0}{a}  \left(\int d^3p\, f - 1 \right) \,.
\end{equation}
This description is valid in the absence of irreducible two-body correlations, which is the case for a smooth matter distribution. The cosmology dependence is encoded in the scale factor $a$, today's background matter density $\rho_0$ and the initial conditions $f_{\rm ini}$.\\
{\it \textbf{Cumulants.}} The cumulants $C^{(n)}$ of the phase space distribution are in practice the physical quantities of interest -- observationally accessible via redshift space distortions and peculiar velocities or numerically determinable from N-body simulations. They encode the number density $n(\v{x})=\exp C^{(0)}$, the velocity $u_i(\v{x})=C^{(1)}_{i}$ and the velocity dispersion  $\sigma_{ij}=C^{(2)}_{ij}$. They can be calculated from $f$ as 
\begin{equation}
\label{cumulants}
G[\v{J}] = \int d^3p\, \exp\left[i\vp\cdot\v{J}\right] f \quad , \quad
C^{(n)}_{i_1 \cdots i_n}:= (-i)^n \left.\frac{\del^n \ln G[\v{J}]}{\del J_{i_1} \ldots \del J_{i_n}} \right|_{\v{J}=0} \,.
\end{equation}
For a general $f$ it is impossible to perform the integration over momentum analytically. Therefore we would like to resort to an ansatz with a specific $\vp$-dependence.

\noindent {\it \textbf{Vlasov hierarchy.}}
The Vlasov hierarchy is constituted by the evolution equations for the cumulants $C^{(n)}$ determined from the Vlasov equation (\ref{VlasovEq})
\begin{equation}
\label{VlasovHierarchy}
\partial_t C^{(n)}  = -\frac{1}{a^2m} \Bigg\{ \nabla \cdot C^{(n+1)}+\sum\nolimits_{|S|=0}^n C^{(n+1-|S|)} \cdot \nabla C^{(|S|)} \Bigg\} - \delta_{n1} m \nabla V \,. 
\end{equation}
It is an infinite coupled hierarchy: in order to determine the time-evolution of the $n$-th cumulant, the $(n+1)$-th is required. The dust model with $C^{(n\geq 2)} = 0$ is the only consistent truncation but fails after shell-crossing where all cumulants become equally important as demonstrated by [\cite{PS09}].

\noindent {\it \textbf{Dust model.}} The dust model is an ansatz for the phase-space distribution function with trivial $\v{p}$-dependence
\begin{equation}
\label{fdust}
f_\d (\vx,\vp,t)= n_\d(\vx,t) \delta\Big(\vp-\vnabla \phi_\d(\vx,t) \Big) \,.
\end{equation}
The cumulants are given by the density $C_{\rm d}^{(0)} = \ln n_{\rm d}(\vx,t) $ and a curl-free velocity $\v{C}^{(1)}_{\mathrm{d}} = \nabla \phi_{\rm d}(\vx,t)/m$ since all higher cumulants vanish identically $C^{(n\geq 2)}_{\mathrm{d}} = 0$. Therefore solving the Vlasov equation for $f_{\rm d}$ is equivalent to solving the coupled fluid system constiting of continuity and Bernoulli equation for $n_{\rm d}$ and $\phi_{\rm d}$.
\begin{equation} 
\label{fluid}
\del_t n_\d = -\,\frac{\vnabla\cdot \left(n_\d \vnabla \phi_\d\right)}{a^2m}\ , \
\del_t \phi_\d = -\, \frac{1}{2}\frac{\left(\vnabla\phi_\d\right)^2}{ a^2 m}  -mV_\d\ , \
\Delta V_\d=\frac{4\pi G\,\rho_0}{a}\Big(n_\d -  1 \Big)  \,.
\end{equation}

\section{The Schr\"odinger method}
 
The Schr\"odinger method (ScM), originally proposed by [\cite{WK93}] as numerical technique to study CDM dynamics, is a special ansatz for the distribution function that is based on the Schr\"odinger Poisson equation 
\begin{equation} 
\label{schrPoissEqFRW}
i\hbar \del_t \psi = - \frac{\hbar^2}{2a^2m} \Delta \psi + m V(\vx)\psi \quad ,\quad 
\Delta V=\frac{4\pi G\,\rho_0}{a}\Bigg(|\psi|^2 -  1 \Bigg) \,.
\end{equation}
If a wavefunction $\psi$ fulfils (\ref{schrPoissEqFRW}) then the Husimi distribution function $f_{\rm H}$ 
\begin{equation} 
\label{Husimi}
f_{\rm H}(\v{x},\v{p},t) = N \left\{  \int d^3y\ \exp\left[-\frac{(\v{x}-\v{y})^2}{4 \sigx^2} - \frac{i}{\hbar} \v{p}\cdot \left(\v{y} -\frac{1}{2}\v{x}\right)\right] \psi(\v{y},t) \right\}^2\,,
\end{equation} 
where $\hbar$ and $\sigx$ are free parameters and $N(\hbar,\sigx)$ is a normalization constant, approximately fulfils the coarse grained Vlasov equation [\cite{T89}] obtained by a Gaussian smoothing of $f$ over $\sigx$ and $\sigp=\hbar/(2 \sigx)$. Therefore, physical processes taking place at scales larger than $\sigx$ and $\sigp$ can be modeled with arbitrary precision.

\noindent{\it \textbf{Key features.}}
The advantages of the Schr\"odinger method compared to the standard dust model are summarized in Tab. 1.  The special $\vp$-dependence of (\ref{Husimi}) allows to compute cumulants analytically. All cumulants are nonzero and can be expressed as Gaussian smoothed functions of $n$ and $\vnabla \phi$ and their derivatives which allows for closing the Vlasov hierarchy, see [\cite{UKH14}].    

\begin{table}[h!]
\label{tab:tabScM}
\centering
\begin{tabular}{|l|c|c|}
\hline
 & Schr\"odinger method $f_{\rm H}(\v{x},\v{p},t) $ & dust model $f_\d(\v{x},\v{p},t) $ \\\hline
 degrees of freedom & 1$\times\mathbb C$: $\psi = \sqrt{n} \exp[i \phi/\hbar ]$ & 2$\times\mathbb R$: $n_\d, \phi_\d$\\
equations of motion & Schr\"odinger-Poisson equation (\ref{schrPoissEqFRW}) & fluid equations (\ref{fluid}) \\
Vlasov equation (\ref{VlasovEq}) solved & approximately ($\hbar,\sigx$) &  exactly\\
shell-crossing  & well-behaved & singularities \\
multi-streaming, virialization & \cmark, \cmark & \xmark\,, \xmark\\
closed-form cumulants & \cmark, $C^{(n\geq 2)} \neq 0$  & (\cmark)\,, $C^{(n\geq 2)}\equiv 0$ \\\hline
\end{tabular}
\caption{Comparison between the Schr\"odinger method and the dust model}
\end{table}

\newpage
\noindent {\it \textbf{Numerical example: Pancake collapse.}}
 In Fig.\,1 we show the standard toy example of plane parallel (or pancake) collapse, whose exact solution in the case of dust is given by the Zel'dovich approximation [\cite{Z70}].  We therefore have analytic expressions for $n_{\rm d}$ and $\phi_{\rm d}$. Nearly cold initial conditions can be implemented by choosing the initial wave function at some early time where shell crossings have not yet occurred as
\begin{equation}\label{inipsi}
\psi_{\rm{ini}}(x)= \sqrt{n_{\rm d}(a_{\rm{ini}},x)} \exp\left[ i \phi_{\rm d}(a_{\rm{ini}},x)/\hbar \right]\,.
\end{equation}

    \begin{figure}[b!]
    \centering
\includegraphics[width=0.90\textwidth]{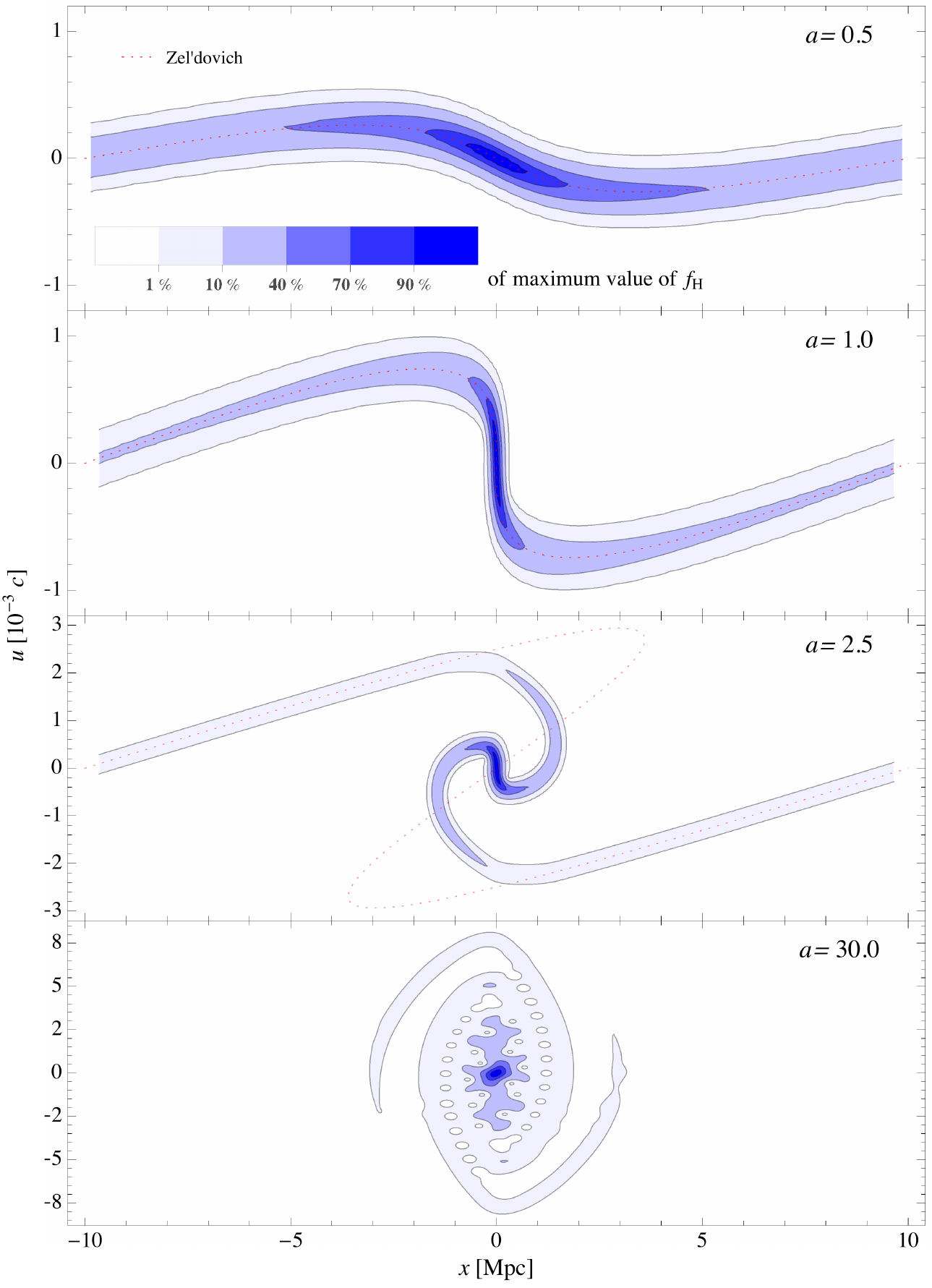}
\caption{\textit{shaded} Schr\"odinger method phase space density $f_{\rm H}$, \textit{dotted} exact dust solution.}
\label{fig:phaseplot}
    \end{figure}
    
    
\section{Coarse-grained dust model}
\noindent The coarse-grained dust model studied in [\cite{UK14}] is limiting case of the Schr\"odinger method when $\hbar\rightarrow 0$ given by
\begin{equation}
\label{fcgdust}
\bar f_\d (\vx,\vp)= \int \frac{d^3x' d^3p'}{(2 \pi\sigx\sigp)^3 } \exp\left[-\frac{(\vx-\vx')^2}{2\sigx^2}-\frac{(\vp-\vp')^2}{2\sigp^2} \right]  
f_\d (\tilde{\vx},\tilde{ \vp})\,.
\end{equation}
It is much closer to the distribution extracted from N-body simulations, which necessarily involves averaging over phase space cells of width $\sigx$ and $\sigp$. Indeed, implementing the coarse-graining in this way results in a resummation in the large scale parameter of the macroscopic model suggested by [\cite{D00}] when the corresponding fluid-type equations are expressed in terms of coarse grained quantities.\\
\vspace{-0.7cm}

  \begin{figure}[h!]
  \label{velspectrum}
 \centering
\begin{tabular}{ll}
\begin{minipage}[l]{0.58\textwidth}
\includegraphics[width=\textwidth]{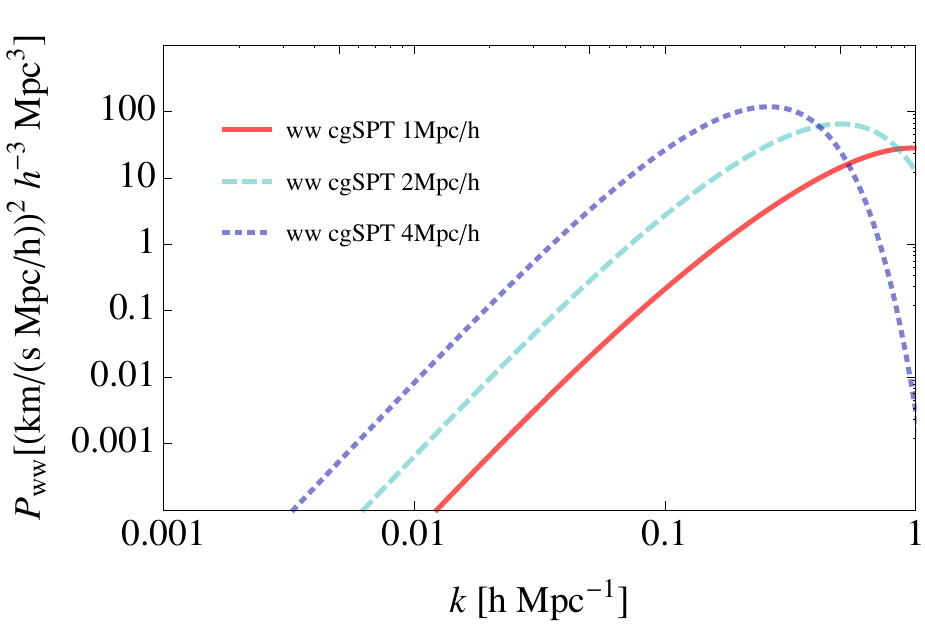} 
\end{minipage} & \begin{minipage}[top]{0.40\textwidth}The coarse-graining naturally leads to a mass-weighted velocity thereby generating large-scale vorticity which is also observed in N-body measurements [\cite{H14}] optimizing the agreement for a smoothing scale of $\sigx=1\,$Mpc. \caption{Power spectrum of vorticity $\v{w}=\vnabla\times\v{v}$ in 1-loop Eulerian perturbation theory for coarse-grained dust (cgSPT) and three different smoothing scales.}\end{minipage}
\end{tabular}
\end{figure}
\vspace{-1cm}
    
    \section{Prospects}
\noindent {\it \textbf{Correlation functions}} of the phase space density are necessary for analyzing observations of large scale structure. Of particular interest is the 2-point correlation function in redshift space $1+\xi(\v{s}) =  \langle (1+\delta(\v{s}_1)) (1+\delta(\v{s}_2)) \rangle$ for biased tracers, like halos or galaxies, relevant to observations made in galaxy surveys. This is investigated for the coarse-grained dust model in [\cite{KUAH14}].

\noindent {\it \textbf{The universality of halo density profiles}} may be understood by determining stationary complex solutions of the Schr\"odinger-Poisson equation. Since the Schr\"odinger method allows for virialization, it could prove useful in further analytical understanding of violent relaxation [\cite{L67}] that leads to universal density profiles [\cite{NFW97}]. These properties might be derived from an entropy principle for collisionless self-gravitating systems as described in [\cite{H12}].
\vspace{-0.5cm}

\newcommand{\apj}{ApJ}
\newcommand{\apjl}{Ap. Lett.}
\newcommand{\apjs}{ApJ Suppl. Ser.}
\newcommand{\mnras}{MNRAS}
\newcommand{\pasj}{PASJ}
\newcommand{\apss}{Ap\&SS}
\newcommand{\aap}{A\&A}
\newcommand{\physrep}{Phys. Rep.}
\newcommand{\mpla}{Mod. Phys. Lett. A}
\newcommand{\jcap}{JCAP}
\newcommand{\prl}{Phys. Rev. Lett.}


\begin{thebibliography}{}



\bibitem[Dominguez, 2000]{D00}
Dominguez, 2000, {\em Phys.Rev.}, D62:103501.

\bibitem[{Hahn}, {Angulo} and {Abel}, 2014]{H14}
{Hahn}, {Angulo} and {Abel}, 2014, arXiv:1404.2280.

\bibitem[{He}, 2012]{H12}
{He}, 2012, {\em \mnras}, 419:1667--1681, 1103.5730.


\bibitem[{Kopp} and {Uhlemann} et~al., 2014]{KUAH14}
{Kopp} and {Uhlemann} et~al., 2014, in preparation.

\bibitem[{Lynden-Bell}, 1967]{L67}
{Lynden-Bell}, 1967, {\em \mnras}, 136:101.

\bibitem[{Melott}, {Pellman} and {Shandarin}, 1994]{M94}
{Melott}, {Pellman} and {Shandarin}, 1994, {\em \mnras}, 269:626, arXiv:astro-ph/9312044.

\bibitem[{Navarro}, {Frenk} and {White}, 1997]{NFW97}
{Navarro}, {Frenk} and {White}, {\em \apj}, 490:493, arXiv:astro-ph/9611107.

\bibitem[Pueblas and Scoccimarro, 2009]{PS09}
Pueblas and Scoccimarro, 2009, {\em Phys.Rev.}, D80:043504, arXiv:0809.4606.

\bibitem[{Takahashi}, 1989]{T89}
{Takahashi}, 1989, {\em Progress of Theoretical Physics Supplement}, 98:109--156.


\bibitem[{Uhlemann} and {Kopp}, 2014]{UK14}
{Uhlemann} and {Kopp}, 2014, arXiv:1407.4810.

\bibitem[{Uhlemann}, {Kopp} and {Haugg}, 2014]{UKH14}
{Uhlemann}, {Kopp} and {Haugg}, 2014, {\em Phys.Rev.}, D90:023517, arXiv:1403.5567.


\bibitem[{Widrow} and {Kaiser}, 1993]{WK93}
{Widrow} and {Kaiser}, 1993, {\em \apjl}, 416:L71.

\bibitem[{Zel'dovich}, 1970]{Z70}
{Zel'dovich}, 1970, {\em \aap}, 5:84--89.

\end{thebibliography}
\end{document}